# Buried spatially-regular array of spectrally ultra-uniform single quantum dots for on-chip scalable quantum optical circuits

*Jiefei Zhang,[1,2] Swarnabha Chattaraj,[3] Qi Huang,[2] Lucas Jordao,[2] Siyuan Lu,[4] and Anupam Madhukar[1,2,3,a]*

[1]Department of Physics and Astronomy, University of Southern California, Los Angeles, California 90089, USA
[2]Mork Family Department of Chemical Engineering and Materials Science, University of Southern California, Los Angeles, California 90089, USA
[3]Ming Hsieh Department of Electrical Engineering, University of Southern California, Los Angeles, California 90089, USA
[4]IBM Thomas J. Watson Research Center, Yorktown Heights, New York, 10598, USA

**Abstract:** A long standing obstacle to realizing highly sought on-chip monolithic solid state quantum optical circuits has been the lack of a starting platform comprising buried (protected) scalable spatially ordered and spectrally uniform arrays of on-demand single photon sources (SPSs). In this paper we report the first realization of such SPS arrays based upon a class of single quantum dots (SQDs) with single photon emission purity > 99.5% and uniformity < 2nm. Such SQD synthesis approach offers rich flexibility in material combinations and thus can cover the emission wavelength regime from long- to mid- to near-infrared to the visible and ultraviolet. The buried array of SQDs naturally lend themselves to the fabrication of quantum optical circuits employing either the well-developed photonic 2D crystal platform or the use of Mie-like collective resonance of all-dielectric building block based *metastructures* designed for directed emission and manipulation of the emitted photons in the horizontal planar architecture inherent to on-chip optical circuits. Finite element method-based simulations of the Mie-resonance based manipulation of the emitted light are presented showing achievement of simultaneous multifunctional manipulation of photons with large spectral bandwidth of ~ 20nm that eases spectral and mode matching. Our combined experimental and simulation findings presented here open the pathway for fabrication and study of on-chip quantum optical circuits.

## I. Introduction

Realization of on-chip scalable quantum optical circuits (QOCs) that allow controlled photon generation and interference to create reconfigurable unitary operations on a photon state with fast (ns to ps) time scale has been a long-sought goal in the field quantum information processing systems [1-4]. A fundamental obstacle has been the absence of a starting on-chip integrable platform of solid state single photon sources (SPSs) that are in regular arrays and spectrally sufficiently uniform. Consequently, exploration of QOCs has been mostly limited to the use of an external laser beam to mimic a non-deterministic SPS for quantum state preparation[5-7]. Epitaxially grown semiconductor quantum dots (QDs) [2,8] and implanted defect-levels [9,10] have been demonstrated to possess many qualities suitable for solid state SPS. In

---

principle, both are integrable with optical elements for controlling SPS emission rate and direction and subsequent manipulation of the emitted photons in an on-chip (i.e. scalable horizontally) architecture that enables controlled interference and entanglement—two phenomena that underpin quantum information processing. However, controlling *simultaneously* the spatial and spectral uniformity adequately for either the dominantly explored semiconductor QD -- the 3D island QD—or defect-based SPSs is still a great challenge [2, 8, 11]. In this paper we report: (a) the realization of a new class of semiconductor mesa top single quantum dots (MTSQDs) [12-14] in arrays (Fig. 1(a)) whose single photon emission purity is ~ 99.5% at 18K and *as-grown* spectral uniformity is < 2nm across a 5×8 array distributed over 1000μm$^2$ (Fig. 1(b)); (b) the embedding of such MTSQD arrays through the overgrowth of a morphology planarizing layer (Fig. 1(d)) to achieve the long-sought platform depicted in Fig.1 (e). Together, this opens the pathway to the fabrication of on-chip integrated QOCs. To this end, we also report here a design for the integration of such buried MTSQDs with the basic light manipulating building units of circuits which can control QD emission rate, direct and guide photon emission utilizing the collective Mie mode of dielectric building blocks (DBBs) [12, 15, 16] as schematically shown as transparent blocks in Fig. 1(f).

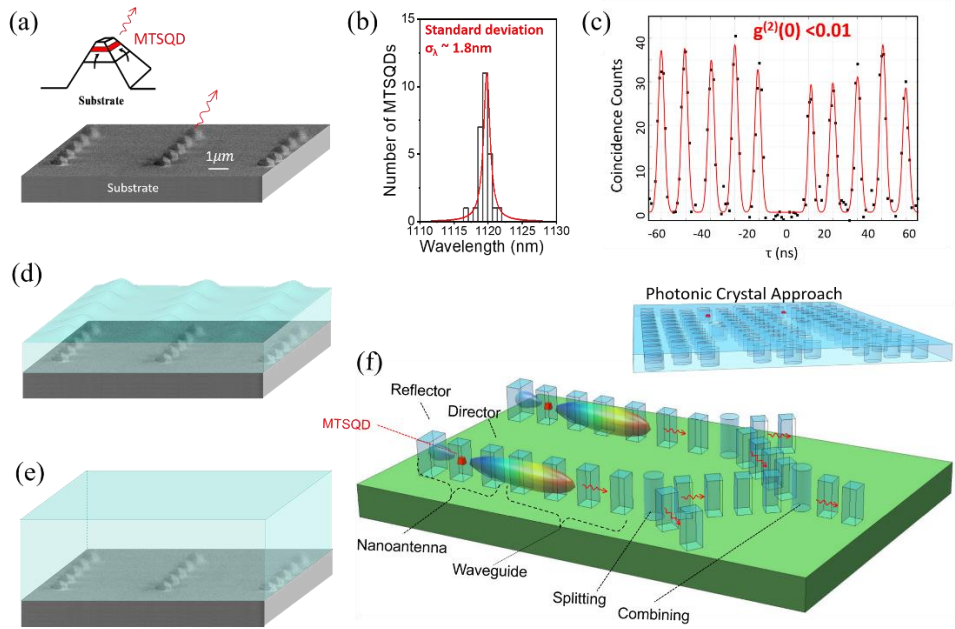

Figure 1. (a) SEM image showing a part of a 5×8 array of spectrally uniform MTSQDs as on-chip scalable single photon sources- that are planarized by overlayer growth (indicated by shaded overlayer)- resulting in a much needed platform for on-chip quantum optical circuits. (b) Histogram of emission wavelengths of a 5×8 array as-grown binary InAs MTSQDs centered at 1120nm. (c) Histogram of correlation counts measured from an InAs MTSQD with 640nm, 80MHz excitation at 18K showing single photon emission with $g^{(2)}(0) < 0.01$. Panels (d) and (e) depict the evolution of the planarization GaAs overlayer growth on the MTSQD arrays such as shown in Panel (a). Panel (f) depicts the two approaches to integration of the needed emitted

photon manipulation elements (cavity and waveguide) implemented via the conventional 2D photonic crystal approach or the new paradigm of exploiting the collective Mie-like resonances[15,16] of all dielectric metastructures. Two SPSs are depicted to emphasize the critical role of panel (e) in enabling creation inter-connected array of multiple SPSs in a horizontal architecture that constitute quantum optical circuits.

## II. Spectrally Ultra-Uniform & Ultra-Pure Single Photon Source Arrays

Figure 1, panel (a) shows a SEM image of part of a 5×8 array of pyramids containing a single quantum dot (SQD) near the mesa top[12]. This unique class of SQD arrays in which the QD locations are controllable to within a few nm is synthesized using size-reducing epitaxial growth on nanomesas. The edge orientations of nanomesas were chosen so that the accompanying surface-curvature stress-gradients direct preferentially the migration of adatoms symmetrically from the sidewalls to the mesa top during growth. This ensures spatially-selective growth on mesa tops [12,14,17]. The approach, dubbed substrate-encoded size-reducing epitaxy (SESRE)[12,14,17,18] enables accommodation of QD forming material combinations with significant lattice mismatch owing to the strain relaxation at the "free" surfaces of the laterally nanoscale mesas, unlike growth in arrays of pits that restrict the material combination to nearly lattice matched [19]. Following such approach, we previously reported arrays of GaAs/In$_{0.5}$Ga$_{0.5}$As/GaAs SQD on the mesa top (MTSQD) exhibiting single photo emission purity ~ 99% ($g^{(2)}(0) < 0.02$)[13]. These InGaAs MTSQDs were centered around 930nm and have a spectral uniformity ($\sigma_\lambda$) of 8nm - a factor of 5 improvement compared to the typically explored lattice-mismatch driven self-assembled quantum dots (SAQDs). More importantly, we showed that there are pairs of *as-grown* MTSQDs emitting within ~300 μeV of each other[13, 16]. Such closely emitting pairs of QDs can be driven into resonance for the study of on-chip photon interference using well established electrical tuning [20,21]. Moreover, these MTSQD SPSs have fine structure splitting (FSS) < 10μeV[13], comparable with the SAQDs and droplet QDs [2, 8, 22, 23], making these MTSQDs suitable for generating entangled photon pairs.

Thus the logical next steps are (1) to further improve on the characteristics of the MTSQDs in terms of spectral uniformity, center wavelength and (2) finding suitable protocols that enable simultaneously the size-reducing growth to form MTSQDs (Fig 1(a)) and the subsequent creation of a morphology planarizing overlayer (Fig.1, panels (d) and (e)) with continued growth. Here we report progress on both these fronts as milestones towards assessing the viability of this class of QDs as on-chip SPS platform needed to realize the long-sought goal of nanophotonic QOCs (Fig.1, panel (f)).

- The Unique Single Quantum Dot Arrays

Our previously reported GaAs/In$_{0.5}$Ga$_{0.5}$As/GaAs MTSQDs, though exhibiting a 5× better spectral uniformity over typical SAQDs, are limited in spectral uniformity largely due to the alloy composition fluctuation induced confining potential fluctuations for each MTSQD in the 5×8 array. One obvious way to improve on the spectral uniformity is to synthesize binary InAs MTSQDs avoiding the inevitable composition fluctuation of alloys. With the SESRE approach, the free surfaces of the nanomesa sidewalls enable relaxation of high lattice-mismatch strain such as 7% for InAs on GaAs(001) for mesa top openings < 50nm without the energetic need for forming strain-relieving defects or surface buckling (i.e. 3D islanding). Indeed, this provides GaAs/InAs/GaAs MTSQDs with flat morphology[19,24]. Moreover, the use of binary InAs as the QD material naturally extends the emission of the MTSQD towards telecommunication wavelengths of 1300nm and potentially up to 1550nm.

The InAs MTSQDs are grown on starting nanomesas of lateral size of ~125nm and depth of ~565nm. A 300 monolayer (ML) GaAs buffer layer with a few monolayer ML thin AlGaAs interspersed is grown at T=600 °C (as measured by pyrometer), $P_{As4}$=2.5E-6 Torr, and Ga delivery time of $\tau_{Ga}$=4 sec/ML (growth rate of 0.25 ML/sec) to (i) recover from any residual damage remaining after deoxidation and (ii) control the mesa top size reduction to bring it to the desired size < 30 nm for the growth of flat single QD on mesa top. The InAs QD is grown at T=480 °C and capped by GaAs to create 3D confinement and to protect the QDs from impurities and defects on the GaAs surface. Details of the substrate patterning, growth conditions, and grown structure can be found in ref. 14. The InAs QDs formed near the apex of the size-reduced nanomesa are estimated to have a base length of ~10-15nm and height ~ 5nm with {101} side walls. Photoluminescence (PL) from each individual MTSQD in the 5×8 array was studied using a home-built micro-PL setup with excitation laser at 640nm and excitation spot focused to 1.25μm for excitation of individual MTSQDs. The emission from the QD is collected by a 40X, NA0.6 objective lens, spectrally filtered by a spectrometer and detected by superconducting nanowire detector. The setup is similar to that previously reported[12, 13] except that the Si avalanching photodiode detectors are replaced by superconducting nanowire detectors for good quantum efficiency at InAs QD emitting wavelength. The measured emission wavelengths from the InAs MTSQDs in a 5×8 array are shown as the histogram in Fig. 1(b). A strikingly narrow spread of 1.8nm centered at 1120nm over an area of ~1000μm$^2$ is found. The observed ultra-uniform emission across the entire array is nearly an order-of-magnitude improvement on the typically reported 3D island QDs [2, 25] and the droplet QDs [26]. It is, to the best of our knowledge, the narrowest spectral emission reported for ordered QD arrays [27-29].

The potential of such highly uniform InAs MTSQDs as near-infrared (NIR) SPSs is confirmed by the second order correlation function of photon emission statistics. The photoluminescence from the InAs MTSQD after spectral filtering was directed into a Hanbury-Brown and Twiss (HBT) setup and subsequently detected by two superconducting nanowire detectors at the transmitted and reflected port of the beam splitter. The measured coincidence count histogram is shown in Fig. 1(c). The data reveal a $g^{(2)}(0) < 0.01$, indicating the single photon emission purity to be > 99.5%. These spectrally ultra-uniform MTSQDs produce ultra-pure single photon emission. This testifies to the potential of the SESRE approach for generating the critically needed spatially-ordered spectrally uniform SPS arrays. To move in the direction of QOC requires, however, planarizing the morphology through appropriate growth of overlayer (Fig.1 (d) and (e)) for the subsequent processing of light manipulating elements. This is reported next.

**III. Morphology Planarizing Overlayer Growth: Towards On-Chip Integration**

For the planarization studies we employed 5×8 arrays of starting square nanomesas fabricated on GaAs(001) substrates using electron beam lithography and wet chemical etching with mesa edges along the <100> direction and lateral sizes in the range of 50nm to 600nm. Two different sidewall configurations were employed: (1) vertical sidewalls made of {100} planes[12,14] and depth of ~185nm; (2) sidewalls of vertical {100} planes to a depth of ~65nm and contiguous {101} planes of equivalent depth of ~65nm (Fig. 2(a)). We dub the latter as mesas on pedestal. The range of the lateral size examined in the same growth enables a study of the evolution of the planarizing overlayer growth as a function of sidewall configuration and depth without the ambiguity of reproducibility of the growth conditions for the MTSQDs and the subsequent morphology planarizing overlayer for all starting nanomesa sizes. In the findings reported below, a total of ~1325MLs of GaAs including a few intervening thin AlAs layers were grown at a pyrometer temperature of $T_{pryo}$~609°C and As$_4$ flux of $P_{As4}$=1.5E-6 torr. A 4.25ML In$_{0.5}$Ga$_{0.5}$As layer is deposited after deposition of 300ML GaAs and AlAs combined layers to form MTSQDs on mesas. This is followed by deposition of another ~1020MLs to bury the QD and reveal the increasing planarization of the overlayer over mesas of decreasing starting lateral sizes for the two different depths and shapes noted above.

For the case of growth on mesas on pedestal, Fig. 2 panels (b), (c), and (d) show SEM and panels (e), (f), and (g) the corresponding AFM images of the surface morphology over three mesas of decreasing starting size of 280nm, 160nm, and 60nm following the overlayer growth.

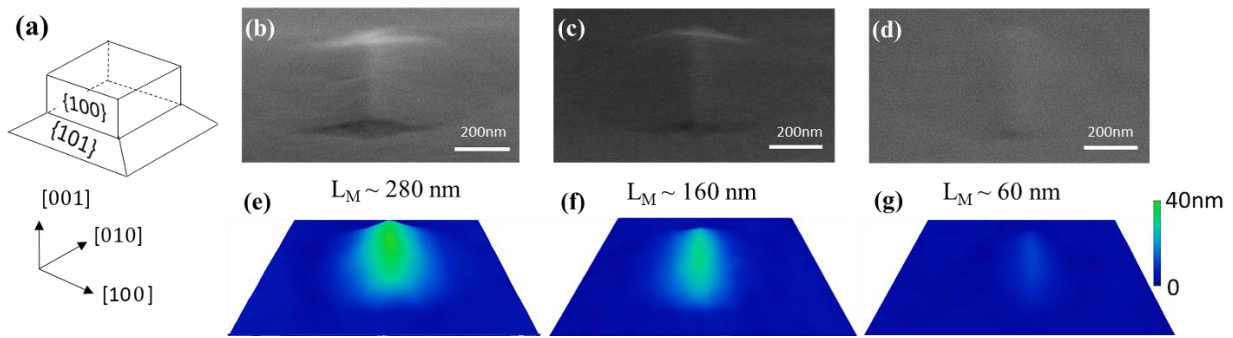

Fig.2. For growth on mesas with pedestal (panel a), shows SEM (panels b, c, and d) and corresponding AFM (panels e, f, and g) images of the surface morphology after ~290nm of a morphology planarization overlayer growth on <100> edge oriented mesas for starting lateral sizes ~280nm, ~160nm, and ~60nm with the vertical {100} sidewalls of depth 65nm and {101} base facets of equivalent depth 65nm.

The various stages of planarization for the same amount of overlayer material delivered (~290nm) under identical growth conditions are evident. Mesas of ~280nm starting size are not completely planarized and reveal an expected ridge type remnant structure[14] with a height of ~20nm and length of 1um along [1 -1 0] direction. With decreasing starting mesa size, the height of the ridge reduces down to <10nm as seen in Fig. 2 (d) and (g), indicating the reducing angle of the side facet of the ridge structure reaching near zero towards planarization. By contrast, for the straight vertical sidewall mesa shapes, the overlayer profile revealed evolution passing through an initial stage of moat-like pits around the base of the mesas that vanish for decreasing starting lateral sizes less than ~300nm. Thus, through choice and control of starting mesa profile that includes appropriate pedestal, depth, and growth conditions, one can simultaneously achieve the required two growth objectives of: (1) formation of the MTSQD at the apex of designed nanomesa lateral size and sidewall profile during mesa-top size-reducing epitaxy (Fig.1 (a)) and (2) subsequent growth of a planarizing overlayer to bury the synthesized MTSQD array in a GaAs matrix (Fig.1 (e)). Such planarized GaAs with MTSQD arrays buried inside provide the starting platform for subsequent integration of light manipulating structures around each MTSQD (Fig.1 (f)), thus enabling fabrication of optical circuits as discussed next.

## IV. Buried SPS Arrays: A Platform for Quantum Optical Circuits

Once suitable buried high-quality ordered arrays of single photon emitting SQDs (Fig.1(b), 1(c) and Fig.2) are available, effectively harvesting and coupling the photons to the light manipulating structures that can be built around each SQD is the essential next step. Thus we have in parallel carried out simulations of the harnessing and manipulation of the photons emitted from such arrays in on-chip horizontal architectures utilizing subwavelength-size DBB of high refractive index (~3.5 for GaAs).

- **Horizontal Architecture for Quantum Optical Circuits:**

Hitherto, the widely explored on-chip integration of QD-SPS with light manipulating elements such as cavity, waveguide, etc. in a horizontal (planar) architecture has been integration with 2D photonic crystal structures[2, 30, 31]. These exploit departures from Bragg scattering of light by array of holes in a photonic crystal membrane[2, 30] to create localized modes that provide the desired functionality. However, as the spectral response of a photonic crystal is typically very narrow (sub-nanometer scale) it presents a formidable challenge in spectral and spatial mode matching between the required light manipulating elements such as the cavity, waveguide and beam-splitter, beam-combiner, the latter two having not yet been implemented to the best of our knowledge. Needed is an integration approach that realizes simultaneously the needed light manipulating functions without the difficulty of mode matching of each functional component and has a wide enough spectral response range (bandwidth) for integration with QDs as well.

Previously we have analyzed the potential of collective Mie-like resonances in all dielectric metastructures made of subwavelength-sized DBB arrays and demonstrated that a single collective spectral mode does provide-- in different spatial regions of the DBB metastrucutre-- the needed functions including enhancing the SPS emission rate, directing the mission, guiding the photons, beam splitting, and beam combining[12,15,16]. This eases the difficulty of mode-matching between different regions ("functional components") of the optical network. Below we provide some illustrative simulated results.

Figure 3(a) shows an illustrative case of a nanoantenna-waveguide metastructure, designed for the ~1120 nm emission of the GaAs/InAs/GaAs QDs reported in Fig.1(b). The MTSQD is embedded in a cubic DBB of size 254nm which is a part of a Yagi-Uda antenna[16]. The larger DBB to the left is the reflector (size 254nm×288nm×254nm) and to the immediate right is the director (cube size 254nm) of the Yagi-Uda antenna. The antenna thus directs photons (red wiggle) towards the continuing DBB chain on the right which acts as a waveguide.

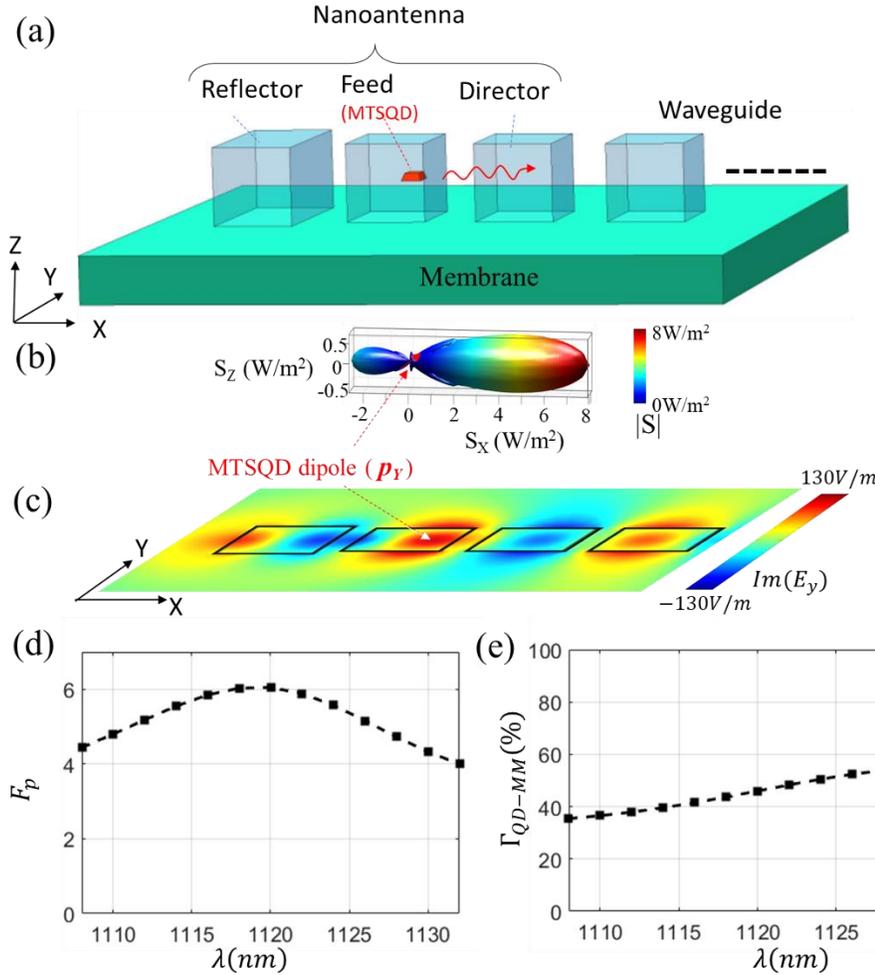

Fig. 3 (a) Yagi-Uda nanoantenna-waveguide structure. The MTSQD is embedded in a cubic DBB of size 254nm. The reflector DBB of the nanoantenna is of size 254nm×288nm×254nm, where as the director DBB and the waveguide DBBs are of cubic shape of size 254nm. (b) Angular distribution of the photon flux on a spherical surface (radius ~ 480nm) surrounding the nanoantenna for a radiating dipole of strength 1-debye at 1120nm. The asymmetry in this angular distribution indicates the nanoantenna effect. (c) Finite element method-based calculated spatial distribution in the nanoantenna-waveguide of the electric field component 90° out-of-phase with the 1-debye point oscillating electric dipole source representing the MTSQD emitting at 1120nm. (d) Purcell enhancement as a function of wavelength, and (e) The fraction of the QD photons that couple to the Mie mode (MM) of the nanoantenna-waveguide structure ($\Gamma_{QD-MM}$) as a function of wavelength. A Purcell enhancement of ~5 over a broad range of ~10nm and $\Gamma_{QD-MM}$ ~50% near 1120nm is indicated.

For this functional metastructure we have used the finite element method to simulate the behavior of the QD emitted photons. The details of the methodology can be found in ref.16. The material system GaAs is mimicked by taking the refractive index of the DBBs as 3.5. Figure 3(b) shows the Pointing vector representing the angular distribution of the photon flux at 1120nm. The Yagi-Uda nanoantenna induced directionality of the flux is evident. Figure 3(c) shows the distribution of the E-field on the XY-plane passing through the center of the DBBs for a MTSQD approximated as a point radiating dipole of strength 1 debye at 1120nm. Note that plotted is the E-field component that is 90° out of phase with the oscillating dipole (Im($E_Y$)) as it is the component responsible for the enhancement of the electric field at the SQD site, the so-called Purcell effect. From the E-field at the location of the QD, we estimate[16] a Purcell enhancement of ~6 around 1120nm over a ~20nm bandwidth as shown in Fig. 3(d). The fraction $\Gamma_{QD-MM}$ of the total emitted photons that couples to the Mie mode of the metastructure we find to be ~ 50% near 1120nm over a ~20nm broad bandwidth as shown in Fig. 3(e). Enabling such large bandwidth in both Purcell enhancement and $\Gamma_{QD-MM}$ is a desirable feature of the Mie-resonance approach. It eases deterministic integration of the MTSQDs with the light manipulating unit as it does away with the requirement of mode matching between different network components (cavity, waveguide, etc.) faced currently in the 2D photonic crystal approach in realizing quantum optical circuits.

The nanoantenna-waveguide unit modelled here is a stepping-stone to creating more complex MTSQD-DBB based optical light manipulating units (LMUs)–with multiple MTSQD SPSs coupled to the collective Mie resonance of the optical metastructure[16]. Coupling multiple SPSs deterministically to the optical circuit created by the inter-connected network of LMUs paves the way to create on-chip entangled states involving photons emitted by different and known MTSQD SPSs. Such DBB based nanoantenna-waveguide, as well as more complex, structures[16] that include beam-splitter, beam-combiner, etc. can be fabricated using electron beam patterning and dry etching on a starting substrate with buried MTSQDs array inside the GaAs matrix (Fig.1 (d) and Fig.2 (g)).

## V. Conclusions

In this paper we have demonstrated two needed milestones towards realizing the basic platform of buried spatially ordered and spectrally uniform single photon sources built upon a unique class of semiconductor quantum dots—the mesa-top single quantum dots (MTSQDs). First, we demonstrated the realization of a 5×8 array of binary InAs MTSQDs with remarkable emission spectral uniformity of 1.8nm over an area of ~1000μm$^2$ with emission centered at 1120nm. Such ultra-uniform MTSQDs are also found to be ultra-pure single photon emitters with > 99.5% purity at 18K revealed by the measured g$^{(2)}$(0) < 0.01. Second, we have demonstrated how to incorporate these QD arrays into a planarized buried configuration, the essential physical platform needed for the subsequent fabrication of light manipulation structures. The significance of the starting mesa profile in controlling the growth front profile evolution at various stages of epitaxial growth is demonstrated. Together, the two now provide the path to the long-sought starting platform for fabricating on-chip quantum information optical circuits.

To this end, of the two approaches to on-chip incorporation of co-designed light manipulation structures -- the 2D photonic crystal or the newly proposed approach of exploiting the collective Mie-like resonance of all-dielectric metastructures – we have presented here simulated response for the 1120nm center emission wavelength of the ultra-narrow spectral emission of the binary InAs MTSQDs. The MTSQDs are embedded in a metastructure that mimics a Yagi-Uda antenna – waveguide combination and thus provides the functions of SQD emission rate enhancement (Purcell effect), directing the emission (antenna effect), and state-preserving guidance of the photons to the desired destination. A Purcell enhancement of ~6 combined with a broad bandwidth of ~20nm is demonstrated – a trade-off of significance in circumnavigating the long standing problem of "mode-matching" faced in integrating individual functions provided by individual discrete device such as resonant cavity, waveguide, beam-splitter, etc.. The ease of achieving spectral and spatial matching between the MTSQD SPS emission and the Mie-like collective mode-based light manipulating metastructure (i.e. the optical circuit), as well as the advantage of having all the network functions being mode-matched, enables design and fabrication of more complex and scalable quantum optical networks. The results presented encourage further exploration of such MTSQDs and their integration for creating and testing quantum optical networks.

In closing we note that although the specific results presented here are for material combinations within the AlGaInAs system the approach to MTSQD array fabrication and their subsequent planarization demonstrated here are general and applicable to a wide range of material combinations. Engineering the MTSQD material combination (III-V arsenides, nitrides, antimonides, etc.), size, and shape, the emission wavelength can be tailored not only for fiber-based optical communication at 1300nm and 1550nm but over the wide range from long- to mid- to near-infrared to visible and ultraviolet regimes for applications ranging from quantum communication, sensing and metrology, to environmental monitoring, and health.

## Acknowledgement

This work was supported by US Army Research Office (ARO), Grant# W911NF-19-1-0025 and Air Force Office of Scientific Research (AFOSR), Grant# FA9550-17-01-0353.

## References

1. E. Knill, R. Laflamme, and G. J. Milburn, Nature **409**, 46-52 (2001).

2. P. Lodahl, S. Mahmoodian and S. Stobbe, Rev. Mod. Phys. **87**, 347 (2015).

3. J. L. O'Brien, A. Furusawa, and J. Vuckovic, Nat. Photon **3**, 687-695 (2009).

4. A. Ryou, D. Rosser, A. Saxena, T. Fryett and A. Majumdar, Phys. Rev. B **97**, 235307 (2018).

5. A. Politi, J. C. F. Matthews, J. L. O'Brien, Science **325**, 1221, (2009).

6. J. W. Silverstone, D. Bonneau, J. L. O'Brien, and M. G. Thompson, IEEE Journal of Selected Topics in Quantum Electronics, **22**, 390-402, (2016).

7. R. Santagati, J. W. Silverstone, M. J. Strain, M. Sorel, S. Miki, T. Yamashita, M. Fujiwara, M. Sasaki, H. Teri, M. G. Tanner, C. M. Natarajan, R. H. Hadfield, J. L. O'Brien, and M. G. Thompson, Journal of Optics **19**, 114006, (2017).

8. P. Michler, Quantum Dots for Quantum Information Technologies (Springer, 2017)

9. A. Sipahigil, R. E. Evans, D. D. Sukachev, M. J. Burek, J. Borregaard, M. K. Bhaskar, C. T. Nguyen, J. L. Pacheco, H. A. Atikian, C. Meuwly, R. M. Camacho, F. Jelezko, E. Bielejec, H. Park, M. Luoncar, M. D. Lukin, Science **354**, 847-850 (2016).

10. E. Togan, Y. Chu, A. S. Trifonov, L. Jiang, J. Maze, L. Childress, M. V. G. Dutt, A. S. Sorensen, P. R. Hemmer, A. S. Zibrov and M. D. Lukin, Nature **446**, 730-734 (2010).

11. T. Schröder, M. E. Trusheim, M. Walsh, L. Li, J. Zheng, M. Schukraft, A. Sipahigil, R.E. Evans, D. D. Sukachev, C. T. Nguyen, J. L. Pacheco, R. M. Camacho, E. S. Bielejec, M. D. Lukin and D. Englund, Nat. Commn, **8**, 15376 (2017).

12. J. Zhang, S. Chattaraj, S. Lu, and A. Madhukar, Jour. App. Phys. **120**, 243103 (2016)

13. J. Zhang, S. Chattaraj, S. Lu, A. Madhukar, Appl. Phys. Lett. **114**, 071102 (2019).

14. J. Zhang, Z. Lingley, S. Lu, and A. Madhukar, Jour. Vac. Sc. Tech. B**32**, 02C106 (2014)

15. S. Chattaraj and A. Madhukar, Jour. Opt. Soc. America B **33**, 2414-2555, (2016).

16. S. Chattaraj, J. Zhang, S. Lu, A. Madhukar, IEEE Jour. Quant. Elec. **56**, 1 (2019).

17. E. Pelucchi, V. Dimastrodonato, A. Rudra, K. Leifer, E. Kapon, L. Bethke, P. A. Zestanakis and D. Vvedensky, Phys. Rev. B. **83**, 205409 (2011).

18. A. Madhukar, Thin Solid Films **231**, 8 (1993).

19. A. Konkar, K.C. Rajkumar, Q. Xie, P. Chen, A. Madhukar, H.T. Lin, and D.H. Rich, J. Cryst. Growth **150**, 311 (1995).

20. R. B. Patel, A. J. Bennett, I. Farrer, C. A. Nicoll, D. A. Ritchie, and A. J. Shields, Nature Photon. **4**, 632-635, (2010).

21. J.-H. Kim, C. J. K. Richardson, R. P. Leavitt, and E. Waks, Nano Lett. **16**, 7061-7066, (2016).

22. R. Seguin, A. Schliwa, S. Rodt, K. Pötschke, U. W. Pohl, and D. Bimberg, Phys. Rev. Lett. **95**, 257402 (2005)

23. Y. H. Huo, A. Rastelli and O. G. Schmidt, Appl. Phys. Lett. **102**, 152105 (2013).

24. A. Konkar, A. Madhukar, and P. Chen, MRS Symposium Proc. **380**, 17 (1995).

25. S. Buckley, K. Rivoire, and J. Vučković, Rep. Prog. Phys. **75**, 126503 (2012).

26. R. Keil, M. Zopf, Y. Chen, B. Hofer, J. Zhang, F. Ding and O. G. Schmidt, Nat. Commun. **8**, 15501 (2017)

27. M. Felici, P. Gallo, A. Mohan, B. Dwir, A. Rudra, and E. Kapon, Small. **5**, 938 (2009).

28. B. Rigal, C. Jarlov, P. Gallo, B. Dwir, A. Rudra, M. Calic and E. Kapon, Appl. Phys. Lett. **107**, 141103 (2015).

29. Y. Chen, I. E. Zedeh, K. D. Jöns, A. Fognini, M. E. Reimer, J. Zhang, D. Dalacu, P. J. Poole, F. Ding, V. Zwiller, and O. G. Schmidt, Appl. Phys. Lett. **108**, 182103 (2016).

30. P. Yao, V.S.C. Manga Rao and S. Hughes, Laser Photonics Rev. **4**, 499-516, (2010).

31. K. H. Madsen, S. Ates, J. Liu, A. Javadi, S. M. Albrecht, I. Yeo, S. Stobbe, and P. Lodahl, Phys. Rev. B. **90**, 155303 (2014).